# Was Einstein Wrong on Quantum Physics?

Mani Bhaumik
Department of Physics and Astronomy,
University of California, Los Angeles, CA 90095.USA
E-mail: bhaumik@physics.ucla.edu

Einstein is considered by many as the father of quantum physics in some sense. Yet there is an unshakable view that he was wrong on quantum physics. Although it may be a subject of considerable debate, the core of his allegedly wrong demurral was the insistence on finding an objective reality underlying the manifestly bizarre behavior of quantum objects. The uncanny wave-particle duality of a quantum particle is a prime example. In view of the latest developments, particularly in quantum field theory, objections of Einstein are substantially corroborated. Careful investigation suggests that a travelling quantum particle is a holistic wave packet consisting of an assemblage of irregular disturbances in quantum fields. It acts as a particle because only the totality of all the disturbances in the wave packet yields the energy momentum with the mass of a particle, along with its other conserved quantities such as charge and spin. Thus the wave function representing a particle is not just a fictitious mathematical construct but embodies a reality of nature as asserted by Einstein.

## 1. Introduction

This year we celebrate with much aplomb the centenary of Einstein's unveiling of his ingenious General Theory of Relativity, although its seed was sown in 1905. In the same *Annus Mirabilis*, he also seeded the other seminal breakthrough of the twentieth century: quantum mechanics. He is granted undisputed credit for the theory of relativity, but receives only guarded recognition for his essential contribution to the quantum revolution. In fact, there is a general impression that Einstein lost the debate on quantum physics.

As we honor him for relativity, it is fitting to ask whether the legendary star of relativity was indeed wrong on quantum physics. Einstein was the first physicist to support the veracity of Planck's radical postulate of quanta of energy. Although proposed by him after years of frustration in formulating his radiation law, Planck himself did not seem to believe in their actual existence. Even more than a decade later in 1913, while recommending Einstein to be a member of the Prussian Academy of Sciences, Planck made a patronizing remark [1], "That he may sometimes have missed the target in his speculations, as, for example, in his hypothesis of light-quanta, cannot really be held too much against him, for it is not possible to introduce really new ideas even in the most exact sciences without sometimes taking a risk." Walther Nernst, another signatory to the recommendation called the light quanta, "probably the strangest thing ever thought up." But Einstein daringly peered through the veil.

Essentially, as early as in 1909 in his Salzburg address [2], Einstein had predicted that physics would have to reconcile itself to a duality in which light could be regarded as both wave and particle. And at the first Solvay Conference in 1911, he had declared [3] that "these discontinuities, which we find so distasteful in Planck's theory, seem really to exist in nature."

So, it was in fact Einstein who fostered the innovative notion of the wave –particle duality by asserting the real existence of quanta of radiation or *photons*, which eventually would open the door for him to his sole Nobel Prize for the photoelectric effect. Following his elicitation, young Louis de Broglie in his PH. D. thesis extended the concept to matter particles with crucial and enthusiastic support from Einstein.

To de Broglie's thesis advisor Langevin, the idea of a matter wave looked far-fetched. So, he sent a skeptical note to his friend Einstein requesting that, 'although the thesis is a bit strange, could he see if it was still worth something.' Einstein replied with a glowing recommendation, "Louis de Broglie's work has greatly impressed me. He has lifted a corner of the great veil. In my work I have obtained results that seem to confirm his." Later Einstein admitted to I. I. Rabi that he indeed thought about the equation for matter waves before de Broglie but did not publish it since there was no experimental evidence for it [4]. De Broglie expressed his appreciation [4a] by writing, "As M. Langevin had great regard for Einstein, he counted this opinion greatly, and this changed a bit his opinion with regard to my thesis." Shortly after reading de Broglie's dissertation, Einstein began suggesting to physicists to look in earnest for an evidence of the matter wave. Soon, proof was furnished with the accidental discovery of electron waves by Davisson and Germer in observing a diffraction pattern in a nickel crystal.

In the meantime Schrödinger, "inspired by L. de Broglie … and by brief, yet infinitely far-seeking remarks of A. Einstein" [5], formulated the wave mechanics of quantum physics, which turned out to be equivalent to the rather abstract matrix mechanics devised by Heisenberg at about the same time. Is it then any wonder that eminent Physicists like Leonard Susskind consider Einstein to be the father of quantum Physics in some sense? [6]

Yet, volumes have been written on Einstein's objection to the implications of quantum physics, particularly to the elements of uncertainty, probability, and non-locality associated with it. There is no question that, as a true scientist, Einstein accepted the extraordinary success and the spectacular results of quantum physics. Can we discern, then, from the very extensive debates and discussions, what was the primary concern of Einstein in his objection to the interpretation of quantum physics? While there can be endless deliberations on this point, why not accept Einstein's own pronouncement on the subject? "At the heart of the problem," Einstein said of quantum mechanics, "is not so much the question of causality but the question of realism." [7]

Bohr was content with his postulate of complementarity of wave-particle duality emphasizing there is *no single underlying reality* that is independent of our observation. "It is wrong to think that the task of Physics is to find out how nature is, Bohr declared. Physics concerns what we can say about nature." [8] Einstein derided this pronouncement as an almost a religious delirium. He firmly believed there was an objective "reality" that existed whether or not we could observe it. [9]

Most contemporary physicists part company with Einstein invoking that it would be futile to look for reality, which gets totally obscure under the thick smoke of the heavy artillery of Hilbert space necessary to deal with particles in quantum mechanics. It is a daunting task indeed to discern any reality in the thickets of a configuration space! *However, if each single particle comprising the ensemble in Hilbert space can be shown to have an objective reality individually, wouldn't it be reasonable to infer that the ensemble in Hilbert space will also have realism even though one may not be able to decipher it?*

In this article, we present a credible allocution in favor of the existence of a physical reality behind the wave function at the core of quantum physics. This is primarily anchored on the incontrovertible physical evidence that all electrons in the universe are exactly alike. We provide reasonable support to show that the wave function of quantum mechanics is not just a conjured mathematical paradigm, but there is an objective reality underlying it, thus justifying Einstein's primary concern of the "the question of realism."

The answer to the long standing puzzle of why all electrons are exactly identical in all respects, a feature eventually found to be shared by all the other fundamental particles as well, was finally provided by the Quantum Field Theory (QFT) of the Standard model of particle physics constructed by combining Einstein's special theory of relativity with quantum physics, which evolved from his innovative contributions.

QFT has successfully explained almost all experimental observations in particle physics and correctly predicted a wide range of phenomena with impeccable precision. By way of many experiments over the years, the QFT of Standard Model has become recognized as a well-established theory of physics. Although one might argue that the Standard Model accurately describes the phenomena within its domain, it is still incomplete since it does not include gravity, dark matter, dark energy, neutrino oscillations and others. However, because of its astonishing success so far, whatever deeper physics may be necessary for its completion would very likely extend its scope without retracting the current fundamental depiction.

## 2. Nature of Primary reality portrayed by quantum field theory

Quantum field theory has uncovered a fundamental nature of reality, which is radically different from our daily perception. Our customary ambient world is very palpable and physical. But QFT asserts this is not the primary reality. The fundamental particles involved at the underpinning of our daily physical reality are only secondary. They are excitations of their respective underlying quantum fields possessing propagating states of discrete energies, and it is these which constitute the primary reality. For example, an electron is the excitation of the abstract underlying electron quantum field. This holds true for all the fundamental particles, be they boson or fermion. Inherent quantum fluctuations are also a distinct characteristic of a quantum field. Thus, QFT substantiates the profoundly counter intuitive departure from our normal perception of reality to reveal that the foundation of our tangible physical world is something totally abstract, comprising of continuous quantum fields that create discrete excitations we call particles.

By far, the most phenomenal step forward made by QFT is the stunning prediction that the primary ingredient of *everything* in this universe is present in *each element of space time* of this immensely vast universe [10]. These ingredients are the underlying quantum fields. We also realize that the quantum fields are alive with quantum activity. These activities have the unique property of being completely spontaneous and utterly unpredictable as to exactly when a particular event will occur [10]. But even to use a word like 'event' renders this activity in slow motion. In actuality, some of the fluctuations occur at mind-boggling speeds with a typical time period of $10^{-21}$ second or less. In spite of these infinitely dynamic, wild fluctuations, the quantum fields have remained immutable, as evinced by their Lorentz invariance, essentially since the beginning and throughout the entire visible universe encompassing regions, which are too far apart to have any communication even with the speed of light. This is persuasively substantiated by the experimental observation that a fundamental particle such as an electron has exactly the same properties, be its mass, charge or spin, irrespective of when or where the electron has been created, whether in the early universe, through astrophysical processes over the eons or in a laboratory today anywhere in the world. Such a precise match between theory and observation infuses immense confidence on our approach.

## 3. A Quantum Particle in Motion

As elucidated above, an electron represents a propagating discrete quantum of the underlying electron field. In other words, an electron is a quantized wave (or a ripple) of the electron quantum field, which acts as a particle because of its well-defined energy, momentum, and mass, which are conserved fundamentals of the electron. However, such an energy-momentum eigenstate of the field can be expressed as a specific Lorentz covariant superposition of field shapes of the electron field along with all the other quantum fields of the Standard Model of particle physics.

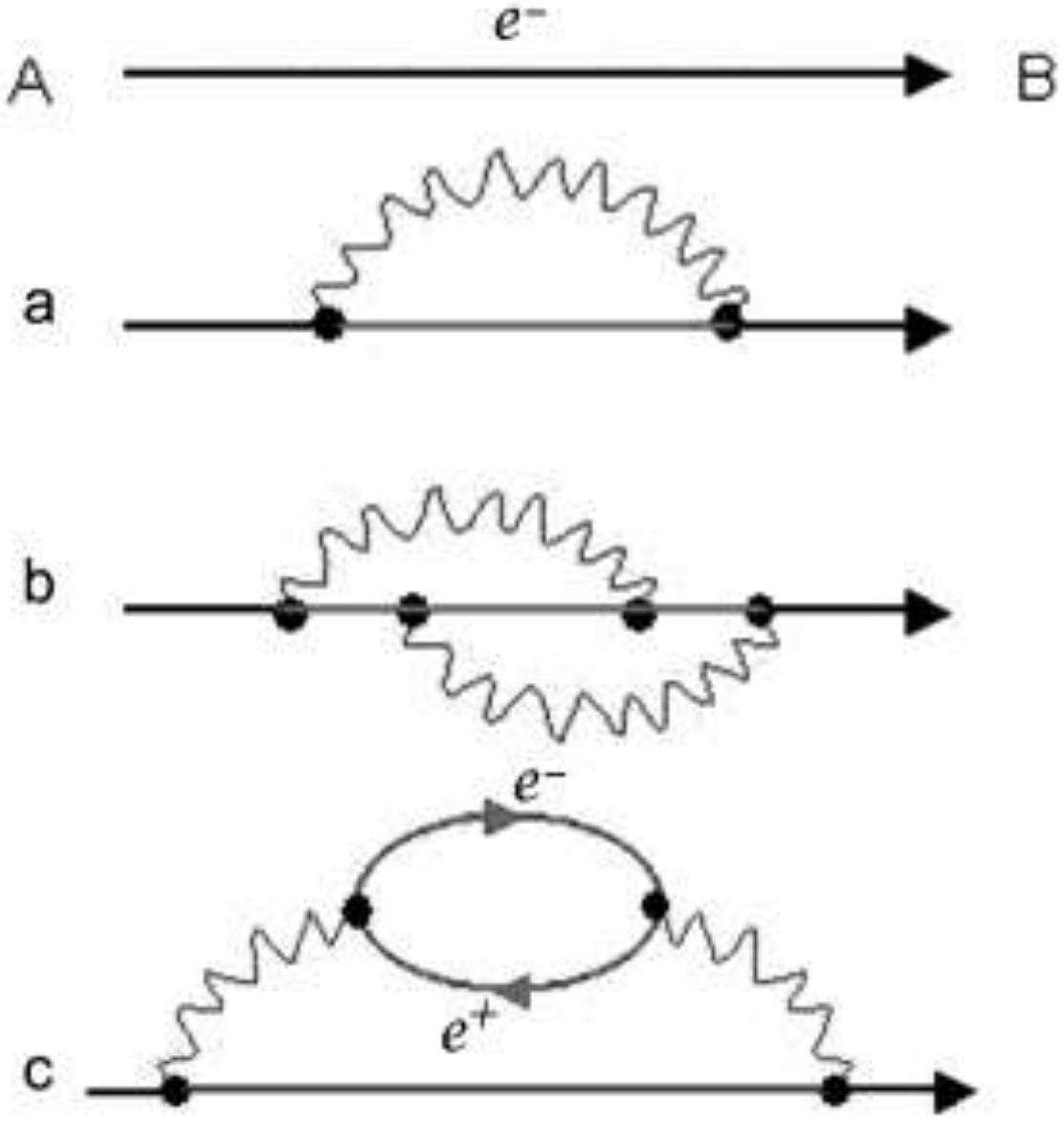

Fig1. Depiction of Feynman diagrams showing some of the various interactions between quantum fields. 1a depicts the interaction of an electron with the photon field, which is commonly described as the emission of a virtual photon by the electron and then reabsorbing it. 1b shows emission of two photons and re-absorption by the electron. The photon in turn can create disturbances in the various quantum fields involving a charge. The virtual photon can emit an electron-positron pair shown in 1c, a muon-anti muon pair, and a quark-antiquark pair and so on.

Superposition of field shapes in a one-particle state evolves in a simple wavelike manner with time dependence $e^{-i\omega t}$. The individual field shapes, each with their own computable dynamic time evolution, are actually the vacuum fluctuations comprising the very structure of the energy-momentum eigenstate. The quantum fluctuations are evanescent in the sense that they pass away soon after coming into being. But new ones are constantly boiling up to establish an equilibrium distribution [11] so stable that their contribution to the electron g-factor results in a measurement accuracy of one part in a trillion [12].

The Lorentz covariant superposition of fluctuations of all the quantum fields in the one-particle quantum state can be conveniently depicted leading to a well behaved smooth wave packet that is everywhere continuous and continuously differentiable.

However, the detailed quantitative calculations are quite involved. For simplicity, we can get the result using a heuristic perspective. For this purpose, let us consider an isolated single quantum of a non-interacting electron quantum field. Since no force is acting on such an electron, its momentum would be constant and therefore its position would be indefinite since a regular ripple from a free electron field with a very well-defined energy and momentum is represented by a delocalized periodic function.

Recalling that the electron in reality is an admixture of all the quantum fields of the standard model, it should be noted that in the non relativistic regime, there would not be enough energy to create any new particle. Consequently, the contribution of the different quantum fields to the single particle would comprise of irregular disturbances of the fields as elegantly depicted by Feynman diagrams (Fig 1), which also aid in calculating the interaction energies amongst the various quantum fields, resulting in the electron ripple to be very highly distorted.

It is well known that such a pulse, no matter how deformed, can be expressed by a Fourier integral with weighted linear combinations of simple periodic wave forms like trigonometric functions. The result would be a wave packet or a wave function that represents a fundamental reality of the universe. Such a wave function would be smooth and continuously differentiable, especially using imaginary numbers in the weighted amplitude coefficients.

Propitiously, it is remarkable to note that a fairly rigorous underpinning of the wave packet for a single particle QFT state in position space for a scalar quantum field, as provided by Robert Klauber [13, eqn. (10-9) pp.275], agrees well with the heuristic approach presented above. The wave function ψ(x) is then given by the Fourier integral,

$$\Psi(\mathrm{x}) = \frac{1}{\sqrt{2\pi}} \int_{-\infty}^{+\infty} \emptyset(k)\, e^{ikx} \mathrm{dk}$$

where $\emptyset(k)$ is a continuous function that determines the amount of each wave number component k = 2π/λ that gets added to the combination.

From Fourier analysis, we also know that the spatial wave function ψ(x) and the wave number function $\emptyset(k)$ are a Fourier transform pair. Therefore we can find the wave number function through the Fourier transform of ψ(x):

$$\emptyset(k) = \frac{1}{\sqrt{2\pi}} \int_{-\infty}^{+\infty} \psi(\mathrm{x})\, e^{-ikx} \mathrm{dx}.$$

Thus the Fourier transform relationship between ψ(x) and $\emptyset(k)$, where x and k are known as conjugate variables, can help us determine the frequency or the wave number content of any spatial wave function.

## 4. The Uncertainty Principle

The Fourier transform correlations between conjugate variable pairs have powerful consequences since these variables obey the uncertainty relation:

$$\Delta x.\ \Delta k\ \geq \frac{1}{2}$$

where $\Delta x$ and $\Delta k$ relate to the standard deviations $\sigma_x$ and $\sigma_k$ of the wave packet. This is a completely general property of a wave packet with a reality of its own and is in fact inherent in the properties of all wave-like systems. It becomes important in quantum mechanics because of de Broglie's introduction of the wave nature of particles by the relationship $p = \hbar k$, where p is the momentum of the particle. Substituting this in the general uncertainty relationship of a wave packet, the intrinsic uncertainty relation in quantum mechanics becomes:

$$\Delta x.\ \Delta p\ \geq \frac{1}{2}\ \hbar$$

This uncertainty relationship has been misunderstood with a rather analogous observer effect, which posits that measurement of certain systems cannot be made without affecting the system. In fact, Heisenberg offered such an observer effect in the quantum domain as a "physical explanation" of quantum uncertainty and hence goes by the name Heisenberg's uncertainty principle. But *the uncertainty principle actually states a fundamental property of quantum systems, and is not a statement about the observational indeterminacy as was emphasized by Heisenberg*. In fact, some recent studies [14] highlight important fundamental difference between uncertainties in quantum systems and the limitation of measurement in quantum mechanics.

Einstein's fundamental objection to the Copenhagen interpretation was its assertion that any underlying reality of the uncertainties was irrelevant and should be accepted under the veil of complementarity. We have established that there indeed is an underlying reality of uncertainty governed by the wave behavior and it traces its origin back to the wave-particle duality first envisioned by Einstein as a reality.

## 5. Role of Probability in Measurement

Having been an expert on statistical mechanics, Einstein was no stranger to probability. In fact he was not opposed to the probabilistic implication of Quantum Physics. As Pauli reported to Born, "In particular, Einstein does not consider the concept of 'determinism' to be as fundamental as it is frequently held to be (as he told me emphatically many times)… In the same way, he disputes that he uses as criterion for the admissibility of a theory the question: *Is it rigorously deterministic?"* [15] As always, he was essentially searching for realism behind the probabilistic outcome in quantum physics.

It should now be evident that the random disturbances caused by the inherent quantum fluctuations of the underlying field is the reason for a quantum particle like an electron to be always associated with a wave function. Such a wave function is by no means just a mathematical construct as considered by many physicists. It represents the totality of all the interactions in the various quantum fields caused by the presence of the electron and facilitated by quantum fluctuations. In other words, a quantum particle like an electron in motion is a travelling holistic wave packet consisting of the irregular disturbances of the various quantum fields. It is holistic in the sense that only the *combination of the disturbances* in the electron field together with those in all the other fields *always* maintains a well-defined energy and momentum with an electron mass, since they are conserved quantities for the electron as a particle.

Because a particle like an electron in motion is represented by a wave function, its kinematics cannot be described by the classical equations of motion. Instead, it requires the use of an equation like the Schrödinger equation for a non-relativistic particle,

$$i\hbar \frac{\partial \psi}{\partial t} = -\frac{\hbar^2}{2m} \nabla^2 \psi + V\psi$$

where $V$ is the classical potential and the wave function $\psi$ is normalized:

$$\int_{-\infty}^{+\infty} \psi^* \psi dx = 1 .$$

The function $\Psi$ evolves impeccably in a unitary way. However, when the particle inevitably interacts with a classical device like a measuring apparatus, the wave function undergoes a sudden discontinuous change known as the wave function collapse. Although it is an essential postulate of the Copenhagen interpretation of quantum mechanics, such a phenomenon has been *perplexing* to the physicists for a long time [16]. However, a behavior like this would be a natural consequence of the distinctive nature of a quantum particle described in this article. In support of this notion, the holistic nature of the wave function is presented as evidence. In a measurement, this holistic nature becomes obvious since the appearance of the particle in one place prevents its appearance in any other place.

Contrary to the waves of classical physics, the wave function cannot be sub-divided during a measurement. This is because *specifically the combination of all the disturbances* comprising the wave function possesses a well-defined energy and momentum with the mass of the particle. Consequently, only the totality of the wave function must be taken for detection causing its disappearance everywhere else except where the particle is measured. *This* *inescapable fact could suggest a solution to the well-known measurement paradox.*

It has been indeed very difficult to understand why, after a unitary evolution, the wave function suddenly collapses upon measurement or a similar other reductive interaction. The holistic nature of the wave function described above seems to offer a plausible explanation. Parts of the wave function that might spread to a considerably large distance can also terminate instantaneously by the process involved in a plausible quantum mechanical Einstein-Rosen (ER) bridge [13] and demonstrated in quantum entanglement of a single photon with the quantum vacuum. [17]

*Thus, the profound fundamentals of our universe appear to support the objective reality of the wave function, which represents a natural phenomenon and not just a mathematical construct.* We also observe that while the wave nature predominates as a very highly disturbed ripple of the quantum field before a measurement, the particle aspect becomes paramount upon measurement.

Because of the wave nature of the particle, the position where the wave packet would land is guided by the probability density $|\psi|^2$ given by Born's rule. It is suggestive to note that Born followed Einstein in this regard as he stated [18] in his Nobel lecture, "Again an idea of Einstein's gave me the lead. He had tried to make the duality of particles -light quanta or photons - and waves comprehensible by interpreting the square of the optical wave amplitudes as probability density for the occurrence of photons. This concept could at once be carried over to the $\psi$-function: $|\psi|^2$ ought to represent the probability density for electrons (or other particles)."

Of course, the exact mechanism by which the wave function collapses is still being highly debated. The most popular version envisions entanglement of the wave function with the constituents of the detector, which decoheres very quickly because of the irreversible thermal motion. One of the principal participant in the development of the theory of decohernce, W. Zurek contends [19] that the Born rule can actually be derived from the theory of decoherence rather than being a mere postulate of quantum theory. There is indeed some support for his contention [20].

The Copenhagen interpretation also requires a conscious observer as an essential part of its formalism, which posits that the reality of a quantum particle or system does not exist until a conscious observer take part in its detection thereby causing the wave function to collapse. Einstein objected to this view by his famous question, "Is the moon there when nobody looks?" Although an observer can bring out a particular reality, the fact that the universe, which is quantum at the core, developed to a mature state even before any observer could appear in it, would give support to Einstein's objection. His contention was that an underlying objective reality should always be present irrespective of measurement.

In contrast, the supporters of the Copenhagen interpretation did not feel it was necessary to delve any further than accepting the wave-particle duality and its consequent uncertainty as a principle of complementarity. In view of the nature of reality discussed in this paper, there is no genuine conflict between Einstein's insistence of an underlying reality and the doctrine of complementarity in the Copenhagen interpretation.

## 6. Quantum Entanglement

Much has been said about how wrong was Einstein in the EPR paper, where he attempted to show that quantum mechanics is incomplete needing further elucidation in the future. For two entangled particles separated by a great distance, Einstein believed there could be no immediate effect to the second particle as a result of anything that was done to the first particle since that would violate special relativity. Quantum mechanics predicted otherwise, which he called, “spooky action at a distance.”

Contrary to Einstein’s expectation, all experimental results so far support non-locality of quantum mechanics. Repeated evidences consistently show that when two particles undergo entanglement, whatever happens to one of the particles can instantly affect the other, even if the particles are separated by an arbitrarily large distance!

Has Einstein's dream of an objective reality been shattered by these experiments? Not necessarily. It is hard to imagine Einstein would have given up just yet. He'd still think there is some deeper reality behind this and certainly that is a reasonable possibility.

Experts such as Maldacena and Susskind [21] postulate that ER=EPR implying there is an as yet unknown quantum mechanical version of a classical worm hole that permits quantum entanglement. There is also a possibility that the quantum fluctuations of the fields are themselves entangled facilitating a quantum mechanical ER bridge [13]. So there still could be an element of reality behind quantum entanglement.

In any case, quantum entanglement does not violate causality or special relativity, since no useful signal can be sent using it. So, Einstein still could have the ultimate chuckle even though in contradiction of his expectation some “spooky action at a distance” has been experimentally demonstrated. More so, because in a serendipitous way, the discovery of quantum entanglement has opened up some groundbreaking applications such as quantum cryptography, quantum computing, and quantum teleportation, which have become areas of very active research. As a consequence, the EPR paper has turned out to be a corner-stone in our understanding of quantum physics. Is that too shabby after being branded wrong?

## 7. Acknowledgement

The author wishes to thank Professors Zvi Bern and James Ralston for important discussions.

## 8. References

[1] G. Kirsten and H. Körber, *Physiker über Physiker,* Akademie Verlag, Berlin (1975) page 201. The English translation is from: A. Pais, *Subtle is the Lord,* Oxford University Press, (2008) page 382.

[2] W. Isaacson, *Einstein: his life and universe,* Simon & Schuster Paperbacks. New York, (2008) page 321.

[3] *The Collected Papers of Albert Einstein,* Princeton University Press, Princeton, vol.3, page 421. Citation provided in [2], page 608.

[4] A. Douglas Stone, *Einstein and the Quantum,* Princeton University Press, Princeton, (2013) page 252. [4a] *ibid* page 250.

5] W. Moore, Schrödinger: *Life and Thought.* Cambridge University Press, Cambridge, (1989) page 211

[6] L. Susskind and A. Friedman, *Quantum Mechanics,* Basic Books, New York, (2014) page xi.

[7] A. Einstein to Jerome Rothstein, *Albert Einstein Archive,* folder 22, document 54, May 22, (1950), Citation provided in [2], page 460.

[8] A. Petersen, *The Philosophy of Niels Bohr,* Bulletin of the Atomic Scientists, (Sept.1963): 12, citation provided in [2], page 333.

[9] Citation in Reference [2] page 334.

[10] F. Wilczek, *The Lightness of Being*, Basic Books, New York, (2008) page 74.

[11] F. Wilczek, *Fantastic Realities,* World Scientific Publishing Co. Pte.Ltd., Singapore (2006) page 404.

[12] S.J. Brodsky, et. al. arXiv: hep-ph/0406325 (2004).

[13] R. D. Klauber. *Student Friendly Quantum Field Theory,* Sandtrove Press, Fairfield, Iowa, 2015.

[14] L.A.Rozema et. al, arXiv: 1202.034 (2012), M.Ozawa, arXiv:1507.02010 (2015); J. Erhart et. al. arXiv 1201.1833 (2012).

[15] *The Born-Einstein Letters, 31 March, 1954,* The Macmillan Press, London (1971) page 221.

[16] R. Penrose, *The Road to Reality,* Random House, London, (2004) page 786.

[17] M. Fuwa et. al. Nature Communications, **6,** article no. 6665 (24 March 2015).

[18] M. Born. Nobel Lecture December 11, 1954, page 262

[19] W.H. Zurek, arXiv: quant-ph/0405161 (2004).

[20] M. Schlosshauer and A. Fine, arXiv: quant-ph/0312058 (2004).

[21]J. Maldacena, L. Susskind, arXiv 1306.0533v2 (2013).